# Hybrid Continuous DoA Estimation with Shared-Radius Co-Prime Circular Arrays

Keyvan Aghababaiyan, *Member, IEEE*

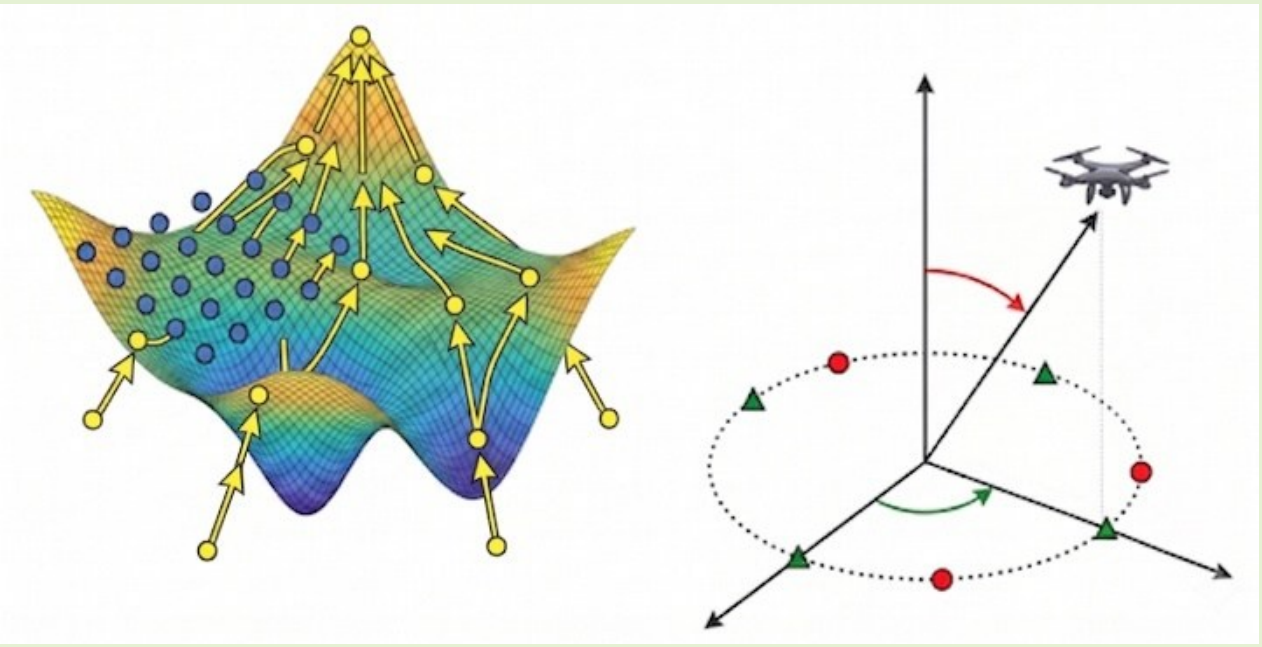

***Abstract*—This paper proposes a shared-radius co-prime circular array for high-resolution, continuous 2D Direction-of-Arrival (DoA) estimation in 3D space, jointly estimating azimuth and elevation angles. The proposed architecture consists of two uniform circular sub-arrays with co-prime antenna counts sharing a common radius $R$R, a design that intrinsically suppresses mutual coupling leakage compared to dense uniform arrays. Unlike existing works that rely on complex phase-mode transformations to map circular structures to virtual linear arrays, we introduce a hybrid continuous-recovery framework operating directly in the physical spatial domain. By integrating a fast, discrete coarse-grid search with a swarm-intelligence continuous refinement stage, the proposed method completely bypasses discrete grid-mismatch limitations and computationally expensive eigenvalue decompositions. A rigorous theoretical analysis using Niven's Theorem establishes the spatial uniqueness of the true source direction, effectively resolving phase ambiguities. Simulation results demonstrate that this hybrid scheme achieves superior resolution and lower Root Mean Square Error (RMSE) at low Signal-to-Noise Ratios (SNR) compared to uniform configurations, while asymptotically converging to the theoretical Cramér-Rao Bound (CRB) at high SNRs.**

## I. Introduction

Direction of Arrival (DoA) estimation is a critical problem in wireless communication systems, particularly in environments with multiple narrow-band signal sources. In systems like massive multiple-input multiple-output (MIMO), DoA estimation is essential for effective downlink precoding and beamforming [1]. Over the years, numerous methods and array structures for DoA estimation have been proposed. Most of these methods focus on uniform array structures due to their simplicity and mathematical tractability. However, the use of non-uniform array structures has been shown to improve the resolution performance of DoA estimation algorithms [2]. To address the limitations of uniform arrays, several non-uniform linear array designs have been introduced, offering greater flexibility and enhanced resolution in DoA estimation [3], [4]. Despite the advantages of non-uniform linear arrays, the nested array structure proposed in [3] is susceptible to mutual coupling effects among closely spaced antennas, which can degrade performance. To mitigate this issue, co-prime arrays were introduced in [4] as a new non-uniform structure, providing a promising alternative that reduces mutual coupling while improving the estimation accuracy. The co-prime linear array proposed in [4] consists of two uniform linear sub-arrays, with $M_1$ and $M_2$ elements, where $M_1$M$_1$ and $M_2$ are co-prime integers. This co-prime linear array structure serves as a foundation for exploring various DoA estimation techniques utilizing co-prime arrays, as discussed in [5]-[7]. In [5], the authors introduced a method based on a total angular field-of-view search, which combines Multiple Signal Classification (MUSIC) applied to the decomposed sub-arrays of the co-prime array. A search-free DoA estimation scheme for co-prime arrays was proposed in [6], based on a projection-like approach. The work in [7] presents the tri-shifted co-prime array which enhances uniform degrees of freedom and spatial efficiency, while reducing mutual coupling effects compared to existing co-prime designs. Although the works in [4]-[7] focus on linear co-prime arrays, they are inherently limited to two-dimensional environments. These schemes cannot be directly applied for DoA estimation in three-dimensional (3D) environments, which require the use of planar arrays to jointly estimate azimuth and elevation angles. A rectangular co-prime array was proposed in [8] to address the challenge of DoA estimation in 3D space using co-prime planar arrays, where large inter-element spacing in each sub-array leads to phase ambiguity. Recently, advanced techniques such as tensor-modeling for EMVS-MIMO radars with sparse geometry [9] and fast 2D-DoA algorithms for conformal MIMO arrays [10] have further highlighted the critical need for high-resolution, hardware-efficient spatial processing. Furthermore, parallel advancements in two-stage reconstruction for co-array estimation [11] and recursive root-finding for multiple parallel sparse arrays [12] continue to drive the demand for robust,

Keyvan Aghababaiyan is a Marie Skłodowska-Curie Postdoctoral Fellow with Universidad Miguel Hernández de Elche, 03202 Elche, Spain (e-mail: kaghababaiyan@umh.es).

interference-free architectures.

In this paper, we propose a shared-radius co-prime circular array for high-resolution, continuous 2D-DoA estimation in 3D space. Unlike the rectangular co-prime arrays in [8], which are limited by planar geometry and susceptibility to phase ambiguities, our circular design provides full 360-degree azimuth coverage and intrinsically suppresses mutual coupling leakage. Furthermore, while recent co-prime circular designs, such as [13], utilize a shared-radius geometry, they typically require Phase Mode Excitation and Bessel function transformations to synthesize virtual linear arrays for subspace-based estimation like ESPRIT. These transformations introduce mathematical approximation errors and heavy computational overhead. In contrast, this paper proposes a novel hybrid continuous sparse-recovery framework operating directly in the physical spatial domain. By integrating a fast, discrete coarse-grid search with a swarm-based continuous refinement stage (PSO), our method completely bypasses beam-space mapping, computationally expensive eigenvalue decomposition (EVD), and the fundamental discrete grid-mismatch limitation. Ultimately, this approach delivers a highly precise, hardware-efficient solution that asymptotically converges to the theoretical CRB, making it ideal for real-time 3D spatial localization.

## II. System Model

In this paper, we present a non-uniform co-prime circular array structure that is formed from two uniform circular sub-arrays with $N_i$Ni, $i \epsilon \{1,2\}$ antenna elements, where $N_1$ and $N_2$ are co-prime integers. To ensure a compact footprint and fair comparison, both sub-arrays share a common radius $R$, leading to inter-element spacings $d_i = 2R\sin\left(\frac{\pi}{N_\mathrm{i}}\right)$ that exceed $\frac{\lambda}{2}$, where $\lambda$ denotes the wavelength of the incident narrow-band signals. The antenna elements of the $i^{th}$ sub-array are located at:

$$L_i = \left\{\left(Rcos\tilde{\theta}_{n,i}, Rsin\tilde{\theta}_{n,i}\right)|\tilde{\theta}_{n,i} = n\frac{2\pi}{N_i},\; 0 \le n \le N_i - 1\right\}. \tag{1}$$

Hence, the locations of the antenna elements in the proposed co-prime circular array are in the set $L = L_1 \cup L_2$. The antenna elements of the two sub-arrays do not overlap (except at the shared physical reference element), and the total number of antenna elements used in the proposed co-prime circular array is obtained as $N_T = N_1 + N_2 - 1$. Fig. 1 shows the schematic of the co-prime circular array when $N_1 = 3$ and $N_2 = 4$.

We assume that there are $M$ uncorrelated narrow-band signal sources from different directions impinging on the array antenna. The $m^{th}$ signal source is located at an elevation angle $\varphi_m$, which is measured downward from the Z-axis, and an azimuth angle $\theta_m$, measured counterclockwise from the X-axis, as shown in Fig. 1. Thus, the received signal vector at the proposed co-prime circular array can be described as [14]:

$$\mathbf{x}(t) = \boldsymbol{A}(\theta,\varphi)\mathbf{s}(t) + \mathbf{n}(t), \tag{2}$$

where $\mathbf{n}(t)$ is a noise vector which contains independent zero-mean unit-variance Gaussian random variables. The term $\mathbf{x}(t)$ can be rewritten as:

$$\mathbf{x}(t) = \sum_{m=1}^{M} \boldsymbol{a}(\theta_m,\varphi_m)\, s_m(t) + \mathbf{n}(t), \tag{3}$$

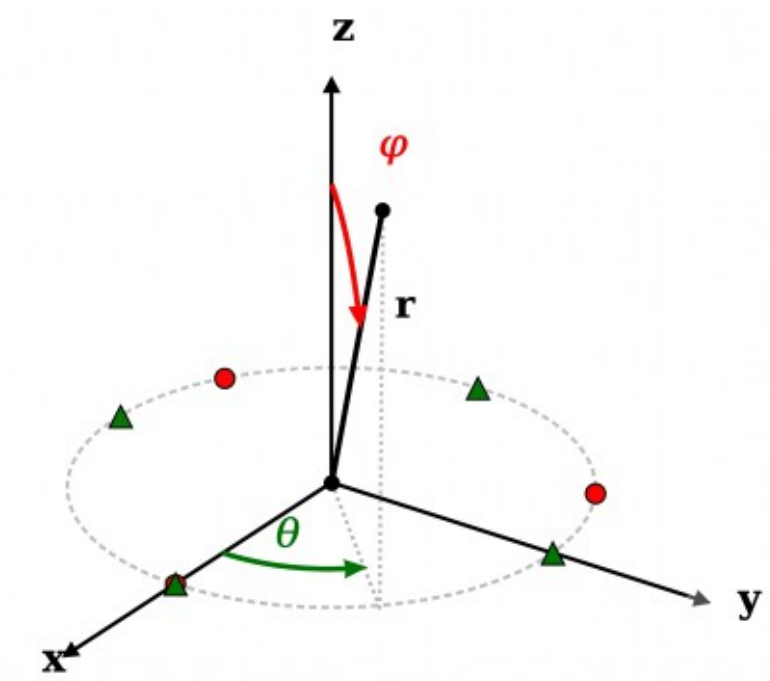

Fig. 1. Geometry of the proposed shared-radius co-prime circular array.

and the matrix $\boldsymbol{A}(\theta,\varphi)$ is the steering matrix, which is defined for different sources as

$$\boldsymbol{A}(\theta,\varphi) = [\boldsymbol{a}(\theta_1,\varphi_1), \boldsymbol{a}(\theta_2,\varphi_2), \ldots., \boldsymbol{a}(\theta_M,\varphi_M)]. \tag{4}$$

In (2), the term $\mathbf{s}(t)$ is the signal sources data vector, and $s_m(t)$ is the signal of the $m^{th}$ source, where $\mathbf{s}(t)$ is defined as:

$$\boldsymbol{s}(\boldsymbol{t}) = [s_1(t), s_2(t), \ldots., s_M(t)]^T. \tag{5}$$

The response of the $n^{th}$ antenna element located at $\tilde{\theta}_n$ corresponding to the $m^{th}$ source is defined by the steering vector element as:

$$a_n(\theta_m,\varphi_m) = exp\left(j\frac{2\pi}{\lambda}R.sin(\varphi_m).cos(\tilde{\theta}_n - \theta_m)\right). \tag{6}$$

Hence, the steering vector $\boldsymbol{a}(\theta_m,\varphi_m)$, i.e., the response of all array elements corresponding to the $m^{th}$ source, is obtained as:

$$\boldsymbol{a}(\theta_m,\varphi_m) = \left[a_1(\theta_m,\varphi_m), \ldots, a_{N_T}(\theta_m,\varphi_m)\right]^T. \tag{7}$$

To resolve the mirror ambiguity inherent in planar arrays, the elevation angle is constrained to the upper hemisphere, i.e., $\varphi_m \epsilon \left[0, \frac{\pi}{2}\right]$.

In practical array implementations, the electromagnetic interaction between closely spaced antenna elements, known as mutual coupling, can severely degrade the DoA estimation performance. For an array with $N_\mathrm{T}$ elements, the received signal model incorporating mutual coupling is modified as $\mathbf{x}(t) = \mathbf{C}\,\boldsymbol{A}(\theta,\varphi)\mathbf{s}(t) + \mathbf{n}(t)$, where $\mathbf{C} \in \mathbb{C}^{N_\mathrm{T}\times N_\mathrm{T}}$ is the mutual coupling matrix (MCM) modeled based on standard impedance interactions [15]. For a circular geometry, the magnitude of the coupling coefficient between the $p^{th}$ and $q^{th}$ elements is inversely proportional to their physical distance $d_{p,q}$. Specifically, the entries of $\mathbf{C}$ are often modeled as $C_{p,q} = c_1\frac{\lambda}{d_{p,q}}exp\left(-j\frac{2\pi}{\lambda}d_{p,q}\right)$ for $p \neq q$, and $C_{p,p} = 1$, where $c_1$ is a complex constant depending on the antenna characteristics.

A major structural advantage of the proposed shared-radius co-prime circular array is its inherent robustness against mutual coupling. In a standard dense Uniform Circular Array (UCA), fitting many elements into a constrained radius forces the inter-element spacing to be less than $\frac{\lambda}{2}$, resulting in large off-diagonal elements in $\mathbf{C}$. In contrast, our co-prime design intrinsically forces the inter-element spacing in each sub-array, given by $d_i = 2R\sin\left(\frac{\pi}{N_i}\right)$, to be strictly greater than $\frac{\lambda}{2}$. Consequently, the off-diagonal coupling coefficients are substantially minimized, making the proposed array geometry highly resilient to mutual coupling errors in practical 3D physical deployments. Although interleaving sub-arrays introduces cross-subarray adjacent elements, the absolute minimum physical distance between any two elements is strictly mathematically bounded by $d_{min} = 2Rsin\left(\frac{\pi}{N_1 N_2}\right)$. By appropriately selecting $R$, $d_{min}$ remains sufficiently large to prevent the deep-coupling failures seen in dense UCAs. To quantitatively validate this structural advantage, we evaluate the mutual coupling leakage energy. The severity of the mutual coupling can be mathematically assessed by measuring the normalized Frobenius norm of the off-diagonal matrix components, defined as:

$$\xi = \frac{\left\|\mathbf{C} - \mathbf{I}_{N_T}\right\|_F}{\left\|\mathbf{I}_{N_T}\right\|_F}, \tag{8}$$

where $\mathbf{I}_{N_T}$ is the $N_T \times N_T$ identity matrix representing an ideal, coupling-free scenario. By substituting the magnitude of the coupling coefficients into the norm, the leakage metric simplifies to:

$$\xi = \frac{1}{\sqrt{N_T}} \sqrt{\sum_{p=1}^{N_T} \sum_{q \neq p} \left| c_1 \frac{\lambda}{d_{p,q}} \right|^2}. \tag{9}$$

In a dense UCA, the condition $d_{p,q} \leq \frac{\lambda}{2}$ causes the squared terms $\left|\frac{\lambda}{d_{p,q}}\right|^2$ to grow significantly, leading to a high leakage energy. In contrast, the spatial constraint $d_i > \frac{\lambda}{2}$ in the proposed co-prime geometry strictly bounds and suppresses these summation terms. As a result, $\xi_{co-prime} \ll \xi_{UCA}$, ensuring that the mutual coupling matrix of the proposed array closely approximates the ideal identity matrix. While this theoretical model effectively captures the first-order distance-dependent geometric leakage, practical hardware implementations also necessitate full-wave electromagnetic calibration to account for complete 3D antenna radiation patterns and platform scattering.

## III. Proposed Continuous DoA Estimation Framework

In this section, we introduce a high-resolution, hybrid continuous sparse recovery framework for 2D-DoA estimation using the proposed co-prime circular array. Traditional discrete methods inherently suffer from the grid-mismatch effect, as the true DoA rarely aligns perfectly with a predefined spatial grid. While increasing the grid resolution can mitigate this error, it exponentially inflates the computational complexity. To overcome this fundamental limitation without relying on computationally expensive EVD or complex phase-mode transformations, we propose a two-stage hybrid approach. Let $\boldsymbol{a}(\theta, \varphi)$ denote the array steering vector for a spatial coordinate.

The first stage executes a fast, discrete correlation-matching process over a highly sparse, coarse 2D dictionary to identify the approximate vicinity of the signal source. Since each atom in this coarse dictionary corresponds to a unique coupled $(\theta, \varphi)$ pair, the algorithm natively avoids the azimuth-elevation pairing ambiguity for multiple sources. The spatial index that maximizes this initial spectrum serves as the initial seed. In the second stage, Particle Swarm Optimization (PSO) is deployed to evaluate the continuous spatial domain exclusively around this seed. By restricting the initial swarm boundary $\delta$ to half the

### Algorithm I:

**Input:** Coarse grid resolution $\Delta_{grid}$= $2^o$, particles $N_p$ = 40, max iterations $T_{max} = 60$.

**Output:** Continuous DoA coordinates $(\theta, \varphi)$.

1: Initialization Phase (Coarse Grid Search):

2: Construct coarse dictionary $Q = \{\boldsymbol{a}(\theta_k, \varphi_k)\}_{k=1}^{K}$, using the spatial steering vector $\boldsymbol{a}$ and grid resolution $\Delta_{grid}$.

3: Evaluate the coarse spatial spectrum:

$$\boldsymbol{P}(\theta_k, \varphi_k) = Re\{\boldsymbol{a}^H(\theta_k, \varphi_k)\mathbf{R}_x\boldsymbol{a}(\theta_k, \varphi_k)\}$$

4: Identify the discrete seed coordinates:

$$(\theta_{seed}, \varphi_{seed}) = argmax\, \boldsymbol{P}(\theta_k, \varphi_k)$$

5: Initialize $N_p$ particles randomly within the continuous cell of the seed, setting the boundary limit $\delta = \frac{\Delta_{grid}}{2}$:

$$\mathbf{X}_i^{(0)} = \left[\theta_i^{(0)}, \varphi_i^{(0)}\right] \sim \boldsymbol{u}(\theta_{seed} \pm \delta, \varphi_{seed} \pm \delta), \quad \forall i \epsilon \{1, \ldots, N_p\}$$

6: Initialize particle velocities $\mathbf{V}_i^{(0)} = 0$, and set personal bests $\mathbf{P}_i = \mathbf{X}_i^{(0)}$.

7: Determine the initial global best $\mathbf{G}_{best}$ based on $\max_i \boldsymbol{P}(\mathbf{X}_i^{(0)})$.

8: Optimization Phase (Continuous Refinement):

9: for $t = 1$ to $T_{max}$ do

10: Update inertia weight $\omega^{(t)}$ using a linear decay mechanism. Linearly decay $\omega^{(t)}$ from 0.9 to 0.05.

11: for $i = 1$ to $N_p$ do

12: Calculate the continuous fitness value:

$$F_i = Re\left\{\boldsymbol{a}^H\left(\mathbf{X}_i^{(t)}\right)\mathbf{R}_x\boldsymbol{a}\left(\mathbf{X}_i^{(t)}\right)\right\}$$

13: if $F_i > \boldsymbol{P}(\mathbf{P}_i)$ then

14: $\mathbf{P}_i \leftarrow \mathbf{X}_i^{(t)}$

15: end if

16: end for

17: Update the swarm's global best: $\mathbf{G}_{best} \leftarrow argmax_{\mathbf{P}_i}\boldsymbol{P}(\mathbf{P}_i)$.

18: for $i = 1$ to $N_p$ do

19: Update velocity using cognitive/social coefficients $(c_1, c_2)$, $c_1 = c_2$=1.49, and uniformly distributed random variables $r_1, r_2 \sim \boldsymbol{u}(0,1)$:

$$\mathbf{V}_i^{(t+1)} = \omega^{(t)}\mathbf{V}_i^{(t)} + c_1 r_1\left(\mathbf{P}_i - \mathbf{X}_i^{(t)}\right) + c_2 r_2\left(\mathbf{G}_{best} - \mathbf{X}_i^{(t)}\right)$$

20: Update the position vector to explore the continuous domain:

$$\mathbf{X}_i^{(t+1)} = \mathbf{X}_i^{(t)} + \mathbf{V}_i^{(t+1)}$$

21: Apply angular boundary constraints to $\mathbf{X}_i^{(t+1)}$.

22: end for

23: end for

24: return Final estimated continuous DoA coordinates: $\left(\hat{\theta}, \hat{\varphi}\right) = \mathbf{G}_{best}$.

coarse grid resolution $\delta = \frac{\Delta_{grid}}{2}$ the particles strictly search the continuous cell containing the peak. The swarm particles dynamically adjust their velocities and positions to maximize the spatial spectrum directly in the physical domain. This continuous refinement completely shatters the discrete error floor, allowing the estimation to tightly converge to the exact continuous DoA coordinates. The complete procedure is summarized in Algorithm I.

## IV. DoA Estimation with Proposed Co-prime Array

DoA estimation schemes for arbitrary array structures are feasible but tend to exhibit high complexity when the array structure is non-uniform. To address this, we propose a method for the proposed co-prime circular array, utilizing the uniform structure of each sub-array and combining their spatial spectra. Consider the two sub-arrays, each a uniform circular array, sharing a common radius $R$ and a physical shared element at $\theta = 0$, as shown in Fig. 1. The steering vector corresponding to the $m^{th}$ signal source for the $i^{th}$ sub-array is defined as:

$$a_i(\theta_m, \varphi_m) = \left[a_{1,i}(\theta_m, \varphi_m), \ldots, a_{N_i,i}(\theta_m, \varphi_m)\right]^T, \quad (10)$$

where $a_{n,i}(\theta_m, \varphi_m) = \exp\left(j.\frac{2\pi R}{\lambda}.\sin(\varphi_m)\cos(\theta_{n,i} - \theta_m)\right)$. The received signal by the $i^{th}$ sub-array is given by:

$$x_i(t) = \sum_{m=1}^{M} a_i(\theta_m, \varphi_m)\, s_m(t) + n(t). \quad (11)$$

When the inter-element spacing is $\frac{\lambda}{2}$, the spatial spectrum $\boldsymbol{P}(\theta, \varphi)$ has a single peak with low resolution and no ambiguity. However, as the shared radius $R$ increases, the spacing beyond $\frac{\lambda}{2}$, $\boldsymbol{P}(\theta, \varphi)$ exhibits multiple sharper peaks. Thus, a single source generates higher resolution at the cost of phase ambiguity. If a signal emitter exists at $(\theta_m, \varphi_m)$, the true phase difference between signals received by two adjacent elements in sub-array $i$ can be derived using the trigonometric identity $\cos A - \cos B = -2\sin\left(\frac{A+B}{2}\right)\sin\left(\frac{A-B}{2}\right)$. Letting $\bar{\theta}_{n,i} = \frac{\theta_{n+1,i}+\theta_{n,i}}{2}$ be the midpoint angle between adjacent elements, the wrapped measured phase difference is:

$$\Delta\varphi_{n,i} = mod\left(-\frac{4\pi R}{\lambda}\sin\left(\frac{\pi}{N_i}\right)sin(\varphi_m)sin(\bar{\theta}_{n,i} - \theta_m), 2\pi\right). \quad (12)$$

The chordal inter-element spacing $d_i$ between two adjacent elements is geometrically related to the radius $R$ by $d_i = 2Rsin\left(\frac{\pi}{N_i}\right)$. By substituting this relation, the relationship between the true unwrapped phase and the measured wrapped phase difference is defined as:

$$\Delta\varphi_{n,i} + 2k_i\pi = -\frac{2\pi d_i}{\lambda}sin(\varphi_m)sin(\bar{\theta}_{n,i} - \theta_m), \quad (13)$$

where $k_i$ is the integer phase ambiguity. To prevent mirror ambiguity across the planar surface, we restrict the elevation to the upper hemisphere, $\varphi_m \epsilon [0, \pi/2]$. Given the constraints $\theta_m \epsilon [0, 2\pi]$, we conclude that $-1 \leq sin(\varphi_m)sin(\bar{\theta}_{n,i} - \theta_m) \leq 1$. This directly provides the bounding range for the ambiguity integer $k_i$:

$$k_i \epsilon \left[-\frac{d_i}{\lambda} - \frac{\Delta\varphi_{n,i}}{2\pi}, \frac{d_i}{\lambda} - \frac{\Delta\varphi_{n,i}}{2\pi}\right]. \quad (14)$$

Because the common radius $R$ is large, the spacing $d_i$ causes $P_i(\theta, \varphi)$ for each sub-array to display multiple grating lobes. To resolve this, we present the following uniqueness theorem. Note that unlike linear co-prime arrays where phase shifts are 1D, our circular geometry introduces a highly non-linear coupled phase term $sin(\varphi_m)sin(\bar{\theta} - \theta_m)$, necessitating a 2D geometric proof.

**Theorem I:** Assume $(\theta_m, \varphi_m)$ is the actual direction of the $m^{th}$ signal source, leading to multiple ambiguous peaks in $P_i(\theta, \varphi)$. By performing joint spatial processing on the spectra of both sub-arrays, the actual direction $(\theta_m, \varphi_m)$ emerges as the strictly unique common peak.

**Existence:** Since $(\theta_m, \varphi_m)$ is the true physical direction of the impinging source, the steering vector naturally perfectly aligns at this coordinate. Consequently, $P_1(\theta, \varphi)$ and $P_2(\theta, \varphi)$ will inherently both exhibit a peak at $(\theta_m, \varphi_m)$.

**Uniqueness:** Assume there are two distinct spatial peaks at directions $(\hat{\theta}_{m,1}, \hat{\varphi}_{m,1})$ and $(\hat{\theta}_{m,2}, \hat{\varphi}_{m,2})$ that appear in the power spectrum of both sub-arrays. Based on (16), the condition for these two directions to yield the same wrapped phase in the $i^{th}$ sub-array requires their spatial deviation to equal an integer multiple of the wavelength:

$$sin(\hat{\varphi}_{m,1})sin(\bar{\theta} - \hat{\theta}_{m,1}) - sin(\hat{\varphi}_{m,2})sin(\bar{\theta} - \hat{\theta}_{m,2}) = \frac{K_i\lambda}{d_i} = \frac{K_i\lambda}{2R\,sin\left(\frac{\pi}{N_i}\right)}. \quad (15)$$

For these ambiguous peaks to perfectly overlap and form a common false peak in both sub-arrays, the spatial deviation must be identical for both geometries. Equating the right side of (18) for $i = 1$ and $i = 2$ yields

$$\frac{K_1\lambda}{2R\,sin(\pi/N_1)} = \frac{K_2\lambda}{2R\,sin(\pi/N_2)}. \quad (16)$$

This simplifies to the ratio:

$$\frac{K_1}{K_2} = \frac{sin(\pi/N_1)}{sin(\pi/N_2)}. \quad (17)$$

According to Niven's Theorem [16], for co-prime integers $N_1, N_2 > 2$ the sines of these rational multiples of $\pi$ are strictly irrational. Thus, any non-zero pair of ambiguity integers $\{K_1, K_2\}$ generates a highly specific discrete irrational constant on the right side of the equation. Since the independent sources in a 3D environment are continuous random variables, the mathematical probability of their spatial deviation perfectly matching this specific measure-zero discrete contour is strictly zero. Consequently, non-overlapping grating lobes systematically fail to coincide, leaving only the true direction $K_1 = K_2 = 0$. Substituting zero into (15) leads to:

$$sin(\hat{\varphi}_{m,1})\,sin(\bar{\theta} - \hat{\theta}_{m,1}) = sin(\hat{\varphi}_{m,2})sin(\bar{\theta} - \hat{\theta}_{m,2}). \quad (18)$$

Because this equality must hold across all adjacent element pairs (various $\bar{\theta}$), we can definitively conclude that $\hat{\theta}_{m,1} = \hat{\theta}_{m,2}$ and $\hat{\varphi}_{m,1} = \hat{\varphi}_{m,2}$. Thus, the true source direction is uniquely identifiable.

Combining Rule (Coincidence Detection): To practically apply **Theorem I**, we implement a coincidence detection rule. Let $\Omega_1$ and $\Omega_2$ be the sets of estimated peak coordinates extracted from $P_1(\theta, \varphi)$ and $P_2(\theta, \varphi)$, respectively. The final

2D-DoA is resolved by finding the intersection of these sets:

$$(\theta_m, \varphi_m) = \text{Arg} \min_{u \in \Omega_1,\ v \in \Omega_2} \|u - v\|_2, \tag{19}$$

where $u$ and $v$ represent the estimated 2D angular coordinate pairs (i.e., $(\theta, \varphi)$) belonging to the peak sets, and $\|.\|_2$ denotes the Euclidean distance. Non-overlapping peaks exceeding the grid resolution tolerance are systematically discarded as grating lobes.

To evaluate the theoretical precision of the proposed shared-radius co-prime circular array, we derive the CRB. For an unbiased estimator, the CRB provides a strict lower bound on the error variance. Note that the CRB is deliberately derived under idealized, coupling-free conditions ($\mathbf{C} = \mathbf{I}_{N_T}$) to establish the absolute theoretical lower bound for the geometric spatial efficiency of the array. Assuming a multivariate Gaussian noise model, the covariance matrix of the received signal is:

$$\mathbf{R}_x = 2\sigma^2 SNR\mathbf{a}(\theta, \varphi)\mathbf{a}(\theta, \varphi)^H + \mathbf{I}. \tag{20}$$

At high SNR or with a large number of snapshots $L$, the Fisher Information Matrix simplifies significantly if the orthogonality conditions $\mathbf{a}^H\dot{\mathbf{a}}_\theta = 0$ and $\mathbf{a}^H\dot{\mathbf{a}}_\varphi = 0$ hold. For our specific geometry, the first derivatives of the array manifold for the $n^{th}$ element in the $i^{th}$ sub-array are:

$$\dot{\mathbf{a}}_\theta(\theta, \varphi) = j\frac{2\pi}{\lambda} Rsin\varphi sin\,(\theta_{n,i} - \theta)\mathbf{a}(\theta, \varphi) \tag{21}$$

$$\dot{\mathbf{a}}_\varphi(\theta, \varphi) = j\frac{2\pi}{\lambda} Rcos\varphi cos\,(\theta_{n,i} - \theta)\mathbf{a}(\theta, \varphi). \tag{22}$$

Evaluating the inner product for the azimuth angle yields:

$$\mathbf{a}^H\dot{\mathbf{a}}_\theta = j\frac{2\pi R}{\lambda} sin\varphi \sum_{i=1}^{2}\sum_{n=0}^{N_i-1} sin(\theta_{n,i} - \theta). \tag{23}$$

Because the elements of each uniform circular sub-array are symmetrically distributed over $2\pi$ (i.e., $\theta_{n,i} = \frac{2\pi n}{N_i}$), the sum of sines and cosines around the full circle strictly evaluates to zero. Thus, the orthogonality conditions are intrinsically satisfied due to the circular symmetry, without requiring mathematical coordinate shifts. Consequently, the CRB equations decouple, and the theoretical bounds are inversely proportional to the squared norms of the derivative vectors:

$$CRB_\theta \approx \frac{1}{4L\sigma^2 SNR|\dot{\mathbf{a}}_\theta|^2} = \frac{1}{4L\sigma^2 SNR\left(\frac{2\pi R}{\lambda}\right)^2 sin^2\varphi.\bar{d}^2}, \tag{24}$$

$$CRB_\varphi \approx \frac{1}{4L\sigma^2 SNR|\dot{\mathbf{a}}_\varphi|^2} = \frac{1}{4L\sigma^2 SNR\left(\frac{2\pi R}{\lambda}\right)^2 cos^2\varphi.\tilde{d}^2}, \tag{25}$$

where the geometric dispersion factors are defined directly by the discrete spatial sampling of the co-prime subsets:

$$\bar{d}^2 = \sum_{i=1}^{2}\sum_{n=0}^{N_i-1} sin^2\,(\theta_{n,i} - \theta), \tag{26}$$

$$\tilde{d}^2 = \sum_{i=1}^{2}\sum_{n=0}^{N_i-1} cos^2\,(\theta_{n,i} - \theta). \tag{27}$$

These expressions mathematically demonstrate that the estimation accuracy of the proposed structure is fundamentally driven by the shared physical radius $R$. By utilizing the co-prime property, the array can employ a strictly larger $R$ without ambiguity, thereby lowering the CRB significantly compared to standard constrained circular arrays.

## V. Evaluation

To validate the performance of the proposed shared-radius co-prime circular array and the hybrid continuous refinement scheme, we conducted extensive numerical simulations. The simulation parameters are set as follows: the coarse search grid resolution is $2^o$ for both azimuth and elevation angles, the number of snapshots is $L = 500$. To guarantee a fair comparison, the normalized circular-array radius is uniformly set to $R = 0.55\lambda$, and the competing Uniform Circular Array (UCA) operates with the exact same physical radius and total number of antenna elements ($N_T = 6$) as the proposed co-prime array. For the continuous swarm optimization stage, the number of particles is set to 40 with a maximum of 60 learning iterations. The PSO parameters ($c_1 = c_2$=1.49, $\omega \in [0.9, 0.05]$) were selected empirically for optimal convergence. Execution times and RMSE are averaged over 30 independent Monte Carlo runs.

Before presenting the numerical results, it is essential to analyze the computational complexity. The complexity of subspace-based methods like 2D-MUSIC is dominated by the EVD of the covariance matrix, which requires $O(N_T^3)$ operations, and the exhaustive 2D fine-grid search, adding $O(N_\theta.N_\varphi.N_T^2)$, where $N_\theta$ and $N_\varphi$ are the number of fine-grid points for azimuth and elevation, and $N_T$ is the total number of antennas. The transformation-based ESPRIT method [13] avoids the 2D grid search but requires complex Bessel function mapping and EVD, yielding a complexity of $O(N_T^3 + M^3)$. In contrast, the proposed hybrid framework evaluates a significantly reduced coarse grid with a complexity of $O(N_{\theta,c}.N_{\varphi,c}.N_T)$, where $N_{\theta,c}$ and $N_{\varphi,c}$ denote the number of search points in the coarse azimuth and elevation grids, respectively. It mathematically refines the estimate via swarm intelligence, requiring only $O(T_{max}.N_p.N_T)$operations, where $T_{max}$ is the maximum number of iterations and $N_p$ represents the total number of swarm particles. By completely bypassing the computationally expensive EVD and exhaustive fine-grid evaluations, and by minimizing $N_T$ via our co-prime design ($N_T = 6$ for $N_1 = 3$, $N_2 = 4$), the computational burden is drastically reduced.

To validate our theoretical complexity analysis, we evaluated the average execution time of the algorithms. We evaluated the average execution time using MATLAB on an Intel Core i5-1335U CPU (1.30 GHz) with 16 GB RAM. The proposed hybrid scheme operates directly in the physical domain and bypasses EVD entirely. Due to the rapid convergence of the swarm refinement stage, it maintains an exceptionally low average execution time of approximately 0.15 seconds. In contrast, the transformation-based ESPRIT [13] requires 0.85 seconds, and the standard fine-grid 2D-MUSIC algorithm takes 1.45 seconds. This confirms that the proposed framework is approximately an order of magnitude faster than 2D-MUSIC and significantly faster than ESPRIT, making it highly efficient

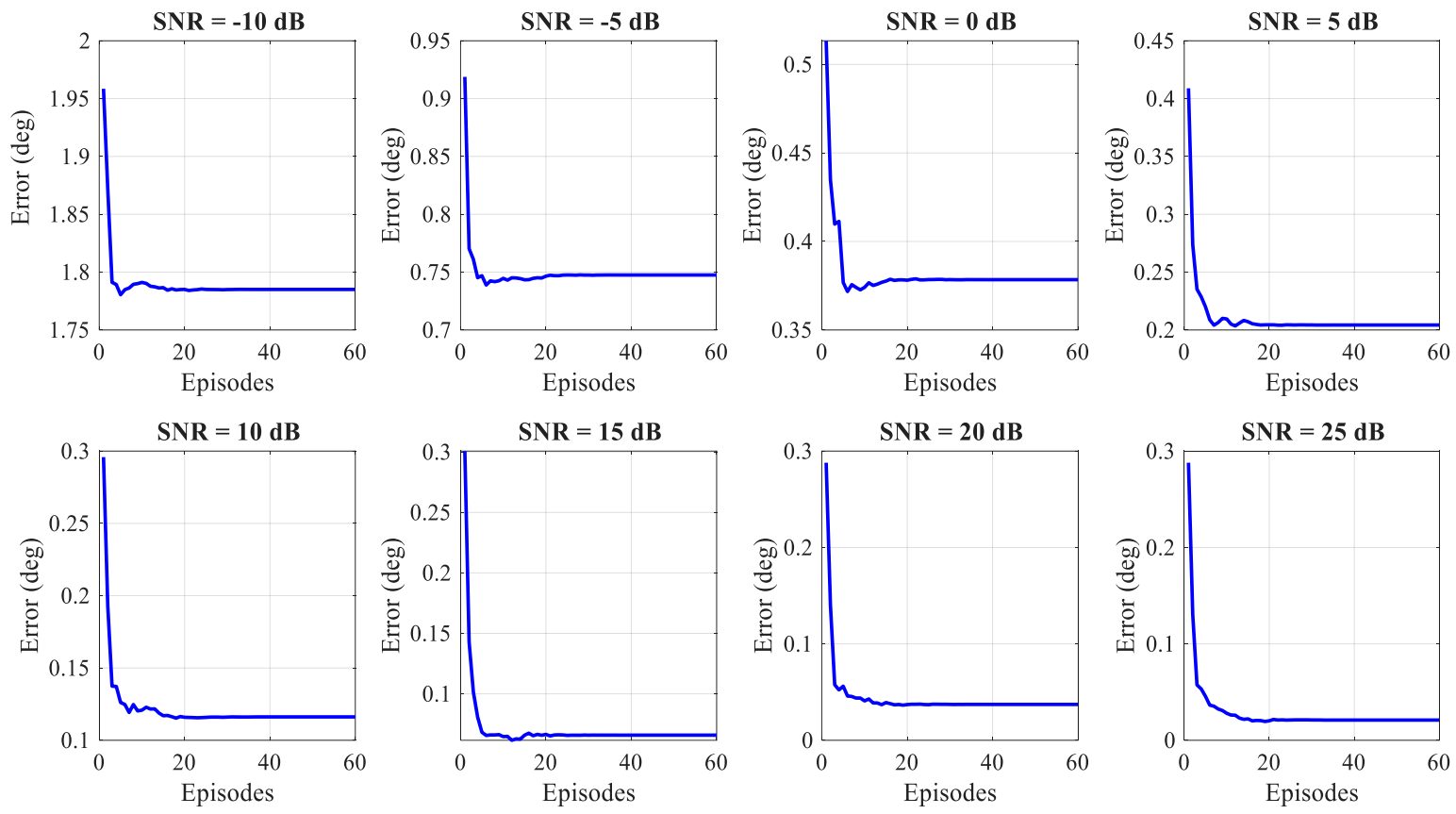

Fig. 2. Convergence behavior of the angle's estimation error over learning episodes across various SNR levels.

and practical for real-time continuous 2D-DoA estimation in 3D space.

To evaluate the computational efficiency and convergence of the proposed continuous swarm-based refinement stage, we analyze its behavior across various Signal-to-Noise Ratio (SNR) regimes. Fig. 2 illustrates the convergence trends for the angles' estimations, measured as absolute estimation error versus the number of learning episodes. The coarse grid resolution of 2° was empirically chosen as an optimal trade-off; a larger grid reduces the initial search cost but negatively impacts the PSO convergence speed. Furthermore, to account for the stochastic nature of swarm initialization, all convergence trends are averaged over 30 independent runs. A critical observation across all SNR levels is the rapid convergence speed. The swarm particles successfully locate the true continuous DoA coordinates within the first 10 to 15 episodes, after which the error sharply drops to a stable steady state. This rapid stabilization dictates that a very small maximum iteration threshold ($T_{max}$) is sufficient to reach optimal accuracy, keeping the computational footprint minimal and making the framework highly suitable for real-time applications. Furthermore, the steady-state residual error exhibits a strong correlation with the SNR. In severe noise conditions (-10 dB), the estimation error stabilizes at approximately 1.78° for azimuth and 2.05° for elevation. However, for high-SNR scenarios (15 dB to 25 dB), the swarm intelligence effectively eliminates the grid-mismatch limitation, driving the steady-state error down to near-zero values (below 0.05°). This robust profile confirms that the continuous refinement stage successfully learns and tracks precise off-grid DoA coordinates without the exhaustive computational burden of fine-grid subspace searches.

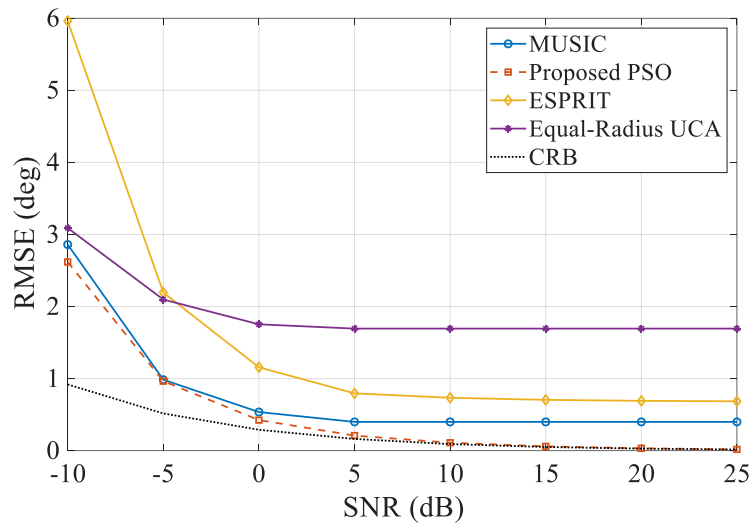

Fig. 3. RMSE performance comparison of the azimuth angle estimation versus SNR for different algorithms.

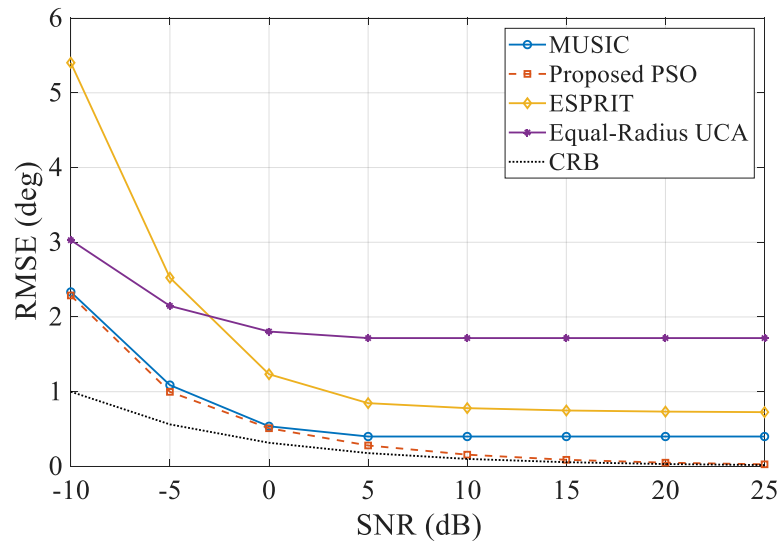

Fig. 4. RMSE performance comparison of the elevation angle estimation versus SNR for different algorithms.

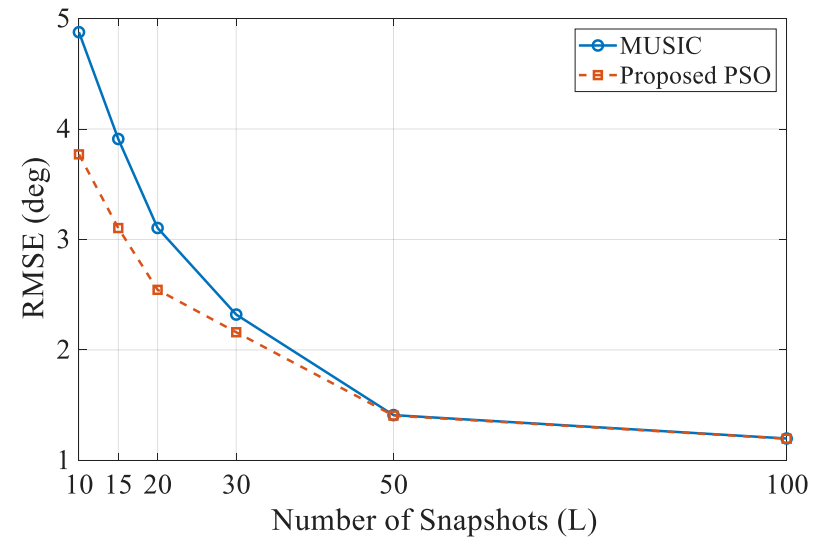

Fig. 5. RMSE performance comparison versus the number of snapshots.

To evaluate the overall estimation accuracy and robustness, the Root Mean Square Error (RMSE) for both angles was analyzed across an SNR range of -10 dB to 25 dB. The proposed hybrid method is benchmarked against standard 2D-MUSIC [13], transformation-based ESPRIT, and an Equal-Radius Uniform Circular Array (UCA) with the identical number of antenna elements. The theoretical CRB is also plotted as the absolute benchmark for statistical efficiency. As depicted in Fig. 3 and Fig. 4, the Equal-Radius UCA exhibits the poorest performance across the entire spectrum. Despite having the same physical footprint and number of elements as the co-prime array, its uniform inter-element spacing significantly exceeds the half-wavelength threshold. This fundamentally violates spatial Nyquist sampling, resulting in unresolvable grating lobes and severe phase ambiguities. Consequently, the UCA's RMSE prematurely plateaus at approximately 1.7°, completely failing to benefit from higher SNR conditions. Furthermore, 3D spatial-spectrum analysis confirms that while standard equal-radius UCAs suffer from severe grating lobes across the elevation-azimuth plane, the proposed co-prime geometry successfully suppresses these spatial ambiguities, yielding a strictly unique spectral peak.

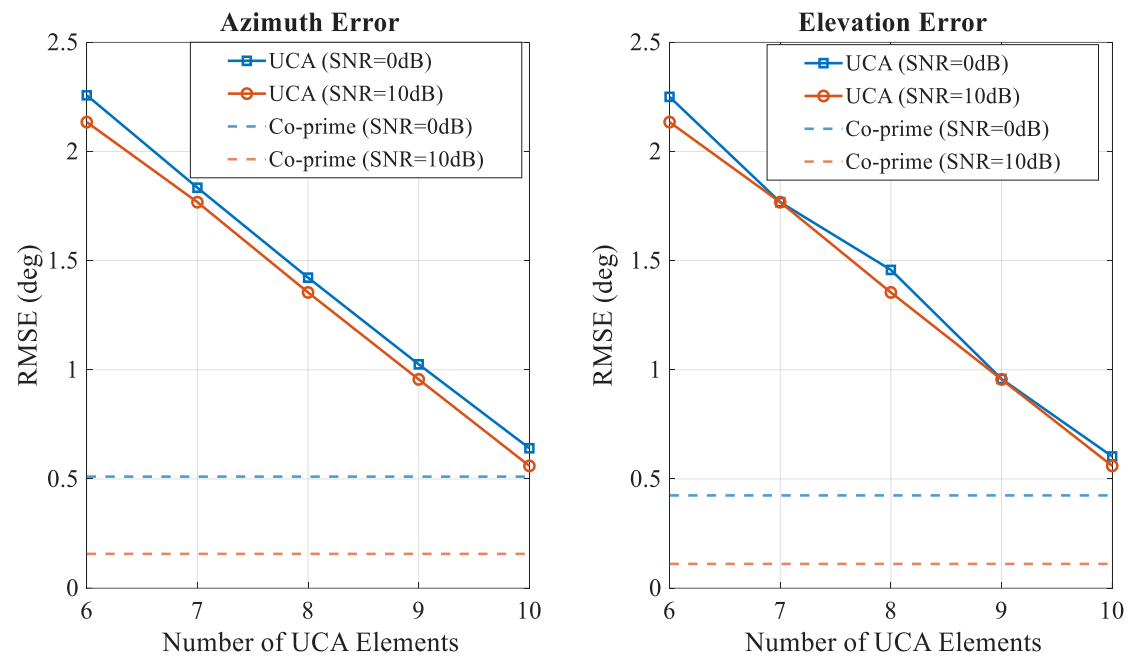

Fig. 6. Impact of increasing the physical antenna count on the estimation error for the baseline UCA compared to the proposed co-prime array.

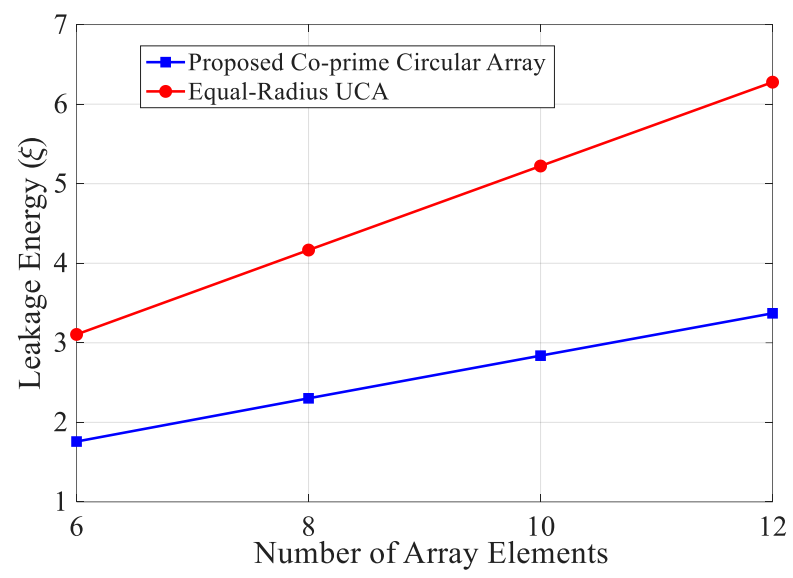

Fig. 7. Mutual coupling leakage energy comparison between the proposed co-prime circular array and the equal-radius UCA as a function of the number of array elements.

The ESPRIT method using co-prime array demonstrates significant vulnerability, particularly in the low-SNR regime (-10 dB), where its error spikes to nearly 6 degrees for elevation and 5.5 degrees for azimuth. Although its accuracy improves as SNR increases, it consistently underperforms compared to MUSIC and the proposed method, eventually plateauing near 0.8 degrees. This performance gap is primarily attributed to the mathematical approximation errors inherent in Phase Mode Excitation and Bessel function transformations required to map the circular array manifold into a virtual linear domain. On the other hand, the 2D-MUSIC algorithm provides stable and reliable estimations, showing a rapid decline in RMSE at lower SNRs. However, at higher SNRs (above 5 dB), the MUSIC curve clearly hits an error floor and plateaus at approximately 0.4 degrees. This flatlining is a direct consequence of the discrete grid-mismatch error. Because standard MUSIC evaluates the spectrum over a fixed, predefined spatial grid, it is mathematically bounded by the grid's resolution and cannot pinpoint the exact off-grid source location, regardless of how clean the signal becomes. In contrast, the proposed hybrid continuous method outperforms all baseline methods. In the harsh low-SNR environment (-10 dB), it achieves the lowest estimation error, showcasing its robust resilience to noise. More importantly, as the SNR increases, the proposed method overcomes the discrete error floor that traps standard MUSIC. Driven by the swarm-based continuous refinement stage, the algorithm successfully navigates the off-grid spatial domain, driving the RMSE down continuously. At high SNRs (15 dB to 25 dB), the proposed method tightly converges toward the theoretical CRB, reaching near-zero error. This asymptotic efficiency mathematically validates that the proposed framework fully resolves co-prime ambiguities while extracting the absolute maximum spatial information available in the physical domain. Furthermore, Fig. 5 evaluates the estimation stability under low-snapshot scenarios (evaluated at SNR = 0 dB). The proposed hybrid method demonstrates robust performance even with as few as $L = 15$ snapshots, significantly outperforming subspace baselines like MUSIC, which inherently suffer from subspace breakdown and require large sample sizes to reliably estimate the covariance matrix.

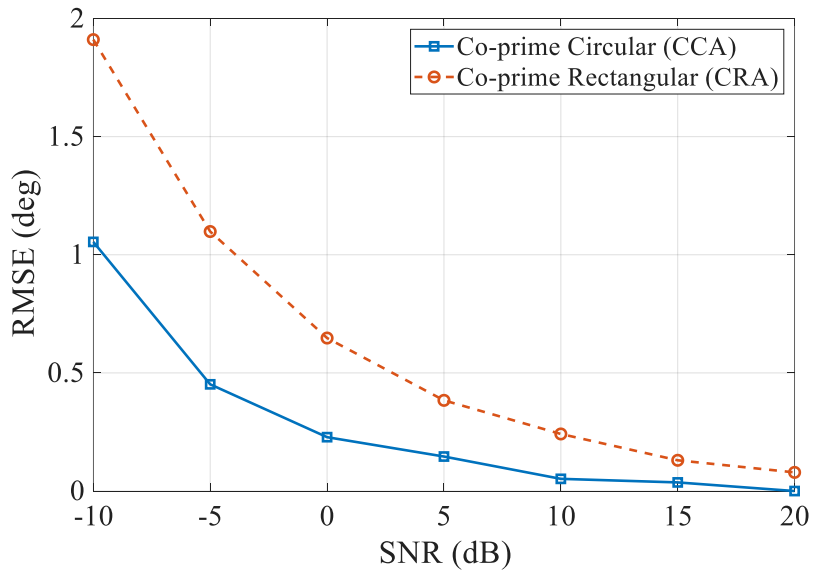

Fig. 8. Azimuth RMSE comparison between the proposed Co-prime Circular Array (CCA) and the Co-prime Rectangular Array (CRA).

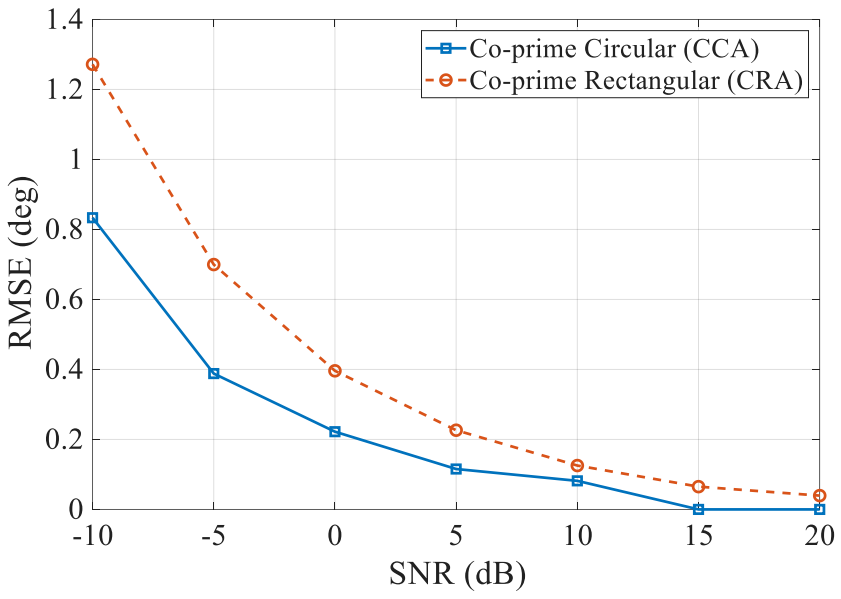

Fig. 9. Elevation RMSE comparison between the proposed Co-prime Circular Array (CCA) and the Co-prime Rectangular Array (CRA).

To further emphasize the hardware efficiency and superior resolving capability of the proposed architecture, we investigate the impact of increasing the physical antenna count in the baseline UCA. Fig. 6 illustrates the azimuth and elevation RMSE of the Equal-Radius UCA as its element count scales from 6 to 10, benchmarked against the constant performance of the proposed 6-element co-prime circular array at moderate (0 dB) and high (10 dB) SNR levels. As expected, densely packing more elements into the constrained UCA geometry gradually improves its estimation accuracy by increasing the spatial sampling density. However, the results reveal a striking structural performance gap. Even when the UCA is heavily populated with 10 antenna elements, its estimation error remains fundamentally worse than that of the proposed co-prime array, which operates with only 6 physical elements. For instance, under a 10 dB SNR scenario, the 10-element UCA struggles to breach the 0.5° RMSE threshold, whereas the proposed co-prime array achieves a highly precise error margin well below 0.2°. This proves that the severe phase ambiguities and grating lobes inherent to uniformly spaced circular arrays cannot be effectively mitigated merely by adding more hardware. By strategically exploiting non-uniform co-prime spacing, the proposed framework inherently synthesizes a larger virtual aperture with enhanced degrees of freedom, thereby delivering significantly superior 3D-DoA estimation accuracy while drastically reducing hardware cost, RF chain complexity, and array footprint.

To quantitatively validate the architecture's robustness against electromagnetic interaction, the mutual coupling leakage energy ($\xi$) is analyzed as a function of the total array elements. Fig. 7 compares the proposed Co-prime Circular Array with a standard Equal-Radius Uniform Circular Array (UCA) as the element count scales from 6 to 12. While densely packing elements into a fixed radius naturally increases mutual coupling, the UCA experiences severe degradation; its leakage energy spikes from approximately 3.1 (6 elements) to over 6.2 (12 elements) as uniform spacing drops below the half-wavelength threshold. Conversely, the proposed co-prime array demonstrates remarkable resilience. By strategically distributing antennas into co-prime subsets, it inherently maintains wider inter-element spacing. Consequently, its leakage energy grows at a substantially slower rate, remaining below 3.4 even with 12 elements.

To evaluate its geometric superiority, the Root Mean Square Error (RMSE) of the proposed 11-element Co-prime Circular Array (CCA) is compared against a 12-element Co-prime Rectangular Array (CRA) across an SNR range of -10 dB to 20 dB in Fig. 8 and Fig. 9. The CCA consistently outperforms the CRA across the entire spectrum despite utilizing fewer hardware elements. This advantage is particularly pronounced in low-SNR environments; for instance, at -10 dB, the CCA yields an azimuth RMSE of approximately 1.05 degrees compared to 1.9 degrees for the CRA, with similar gaps in elevation accuracy. This performance discrepancy stems from the inherent structural symmetry and 360-degree coverage of the circular geometry, which effectively suppresses the phase ambiguities and spatial overlaps that severely degrade flat rectangular arrays. Furthermore, at higher SNRs, the CCA's error converges toward near-zero values at a much steeper rate. These results indicate that the proposed CCA architecture is both more hardware-efficient and significantly more precise for 3D DoA estimation.

## VI. Conclusion

This paper proposed a high-resolution, low-complexity continuous 2D-DoA estimation framework using a shared-radius co-prime circular array. By integrating a coarse physical-domain search with a swarm-intelligence continuous refinement stage (PSO), we successfully resolved phase ambiguities and eliminated discrete grid-mismatch limitations without requiring complex beam-space transformations or computationally expensive eigenvalue decompositions. Mathematical proof via Niven's Theorem confirmed the spatial uniqueness of the true DoA. Furthermore, quantitative analysis validated that the proposed co-prime geometry inherently suppresses mutual coupling leakage compared to dense uniform arrays. Ultimately, the proposed hybrid approach offers a highly efficient alternative that maintains superior accuracy at low SNRs and asymptotically converges to the theoretical Cramér-Rao Bound at high SNRs, making it an ideal, hardware-efficient solution for real-time 3D spatial localization. Despite these structural advantages, practical deployments must also address cross-subarray mutual coupling at the interleaving points, which typically requires full-wave electromagnetic calibration.

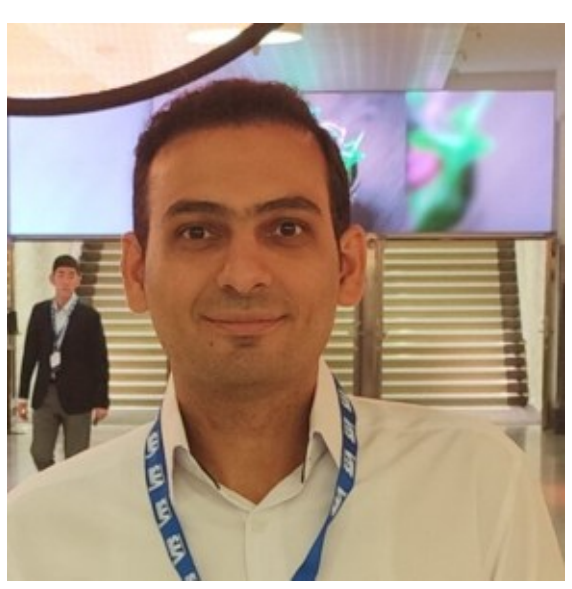

**Keyvan Aghababaiyan** received the B.Sc. degree (Hons.) in Communication Engineering from Amirkabir University of Technology, Tehran, Iran, in 2011, the M.Sc. degree in Communication Engineering from Sharif University of Technology, Tehran, Iran, in 2013, and the Ph.D. degree in Communication Engineering from the University of Tehran, Tehran, Iran, in 2019. During his academic career, he was awarded multiple prestigious fellowships from the Iranian National Elite Foundation, received consecutive Best Ph.D. Student Awards, and graduated as the top-ranked Bachelor's student in his cohort. From 2020 to 2022, he was a Postdoctoral Researcher at the University of Tehran, supported by the Iranian National Elite Foundation. He is currently a Marie Skłodowska-Curie Postdoctoral Fellow at the Networked Systems Lab, Universidad Miguel Hernández de Elche (UMH), Spain. His research interests include deterministic networking, intelligent resource allocation, and next-generation wireless communication systems.